# Electrodeposition of Ni from Various Non-Aqueous Media: The Case of Alcoholic Solutions

K. Neuróhr[1+], L. Pogány[1], B. G. Tóth[1], Á. Révész[2],

I. Bakonyi[1], L. Péter[1]

[1]*Wigner Research Centre for Physics, Hungarian Academy of Science.
1121 Budapest, Konkoly-Thege út 29-33, Hungary*

[2]*Department of Materials Physics, Eötvös University.
H-1117 Budapest, Pázmány Péter sétány 1/A, Hungary*

ABSTRACT − Although electrodeposition from aqueous media has been widely used to obtain metallic deposits, there are cases where the application of non-aqueous solutions offers advantages over the traditional baths or even represents the only way to electrodeposit some metals. For this reason, a study of the electrodeposition of Ni from various alcoholic solutions was performed. Apart from a detailed cyclic voltammetry study of these solutions, the surface morphology, crystal structure and texture as well magnetic properties of the deposits have also been investigated. The best results were obtained with methanol as solvent, so results on Ni deposits prepared from this solution will be presented in more detail whereas the case of the other alcoholic solutions investigated will be summarized only briefly. Structural and magnetic properties of the deposits obtained have been compared to literature data on Ni samples obtained from various non-aqueous solvents.

KEYWORDS: nickel, electrodeposition, non-aqueous media, alcoholic solutions

---

[+] Corresponding author. E-mail: neurohr.katalin@wigner.mta.hu

# Introduction

Electrodeposition of metals and alloys has long been an important topic of scientific and technical literature.[1] The overwhelming majority of technologically relevant electroplating procedures are based on various aqueous solvents.[2] In spite of the scattered appearance of the application of non-aqueous solvents, the breakthrough happened only when ionic liquids became a hot topic and many papers appeared about the electrodeposition of such metals from these electrolytes[3] that are not accessible with conventional aqueous electroplating.

Apparently, Takei initiated first a series of studies for the electrodeposition of metals from non-aqueous solutions.[4-8] The main motivation for his efforts was[7] the need for obtaining electrodeposits of metals like Al and Be which cannot be electrodeposited from aqueous solutions. There are also several other metals or alloys which have useful properties from the viewpoint of industrial applications but cannot be electrodeposited from aqueous solutions. In spite of the efforts after the works by Takei, the literature of metal and alloy electrodeposition from non-aqueous solutions is not very abundant (apart from the extended literature of electrodeposition from ionic liquids which was reviewed in Ref. 3).

As to the electrodeposition of Ni from non-aqueous solutions, several papers have been published on this topic[3,4,6,7,9-16] but we shall here only briefly mention some of these results. Lee et al.[14] reported the growth of nanocrystalline Ni films on n-Si(111) substrate by pulsed electrodeposition. A room-temperature methanol solution with dissolved $NiCl_2$ (in 0.1 M concentration by using the hydrated $NiCl_2$ salt) was used which was purged with high-purity nitrogen to remove the oxygen. The potential pulses were applied at various frequencies with a constant duty cycle of 50%. Nickel was also electrodeposited from a $Ni(CF_3COO)_2$-halide-methanol bath.[6,7] Plating of nickel with various ceramic particles (also referred to as suspension plating) was carried out using an ethanol-based nickel bath. The effects of particle loading, current density and the water content of the bath on the particle codeposition were studied.[15]

There have been also reports on the electrodeposition of Ni alloys from non-aqueous solutions. The codeposition and the effects of the electrodeposition parameters of La-Ni alloy films were investigated in ethanol.[17] Electrodeposited $(NiZn)_{33}Fe_{67}$ and $Ni_{33}Fe_{67}$ alloys were obtained from ethylene glycol as solvent and metal sulfates as solutes.[18-19] Recently, Saranya et al.[16] studied the electrodeposition of Ni, Cu and Ni-Cu alloys from a triethylammonium trifluoroacetate medium.

In the case of an aqueous electrolyte solution, the reduction processes of metal ions generally strongly depend on the pH of the solution. Furthermore, under the application of a high overpotential, many hydrogen bubbles can be generated on the surface of the working electrode, and this can make the coating uneven. The alkalization of the solution in the vicinity of the cathode may lead to the formation of metal-hydroxide precipitates which causes a deterioration of the coating upon incorporation.

A possible way of getting rid of these problems is to carry out electrodeposition from non-aqueous solutions. An advantage is that the anion formation in the solvent decomposition process can not lead to the occurrence of a precipitation with the metal ion. A further benefit of using non-aqueous solvents may be that the metal of interest can be deposited at overpotentials where electrodeposition in aqueous solutions is not possible due to the vigorous hydrogen evolution. In spite of the numerous studies published so far, little is known about the details of the electrodeposition of metals from non-aqueous solutions since most of the studies have been dealing with the deposit quality rather than the fundamentals of the processes.

As the above literature survey shows, the recognition of the prospective benefits of the non-aqueous solvents was not followed by a systematic study of the possible solvents as electrodeposition media but the studies published so far have rather focused on the application of one solvent of choice only. In this study, a comparative study of nickel electrodeposition is presented by using a wide range of alcoholic solvents. Besides methanol and ethanol as solvents, Ni electrodeposition from ethylene glycol, glycerol, 1,2-propanediol and 1,3-propanediol was also the aim of our research.

As to the source of $Ni^{2+}$ ions, we have taken a different approach than followed in previous reports which all used either sulfamate salt of nickel[13] or synthesized an appropriate solute (e.g., $Ni(CF_3COO)_2$, see Refs. 4 and 16) to dissolve in the organic solvent. All these nickel salts previously used as solute are uncommon and expensive chemicals. Nickel salts which are available in anhydrous form were considered for the preparation of the solution. Both the sulfate and the perchlorate salts of nickel are hardly soluble in most of the organic solvents. The Ni ions are divalent, so only monovalent anion salts are applicable because of the solvation. The nitrate is a monovalent anion but nickel(II) nitrate is not available in anhydrous form. We have finally chosen the nickel chloride as solute in the alcoholic solvents. Nickel chloride is commercially available in non-hydrated (dry) form and it is soluble in all alcohols used in the present study.

The fundamental electrochemical description of the chosen electrolyte systems was complemented by a scanning electron microscopy (SEM) investigation of the coating quality, the structural characterization with X-ray diffraction (XRD) and magnetization studies.

## Experimental

*Chemicals and materials.* — The substrate was a Pd-coated polycrystalline Cu foil because preliminary experiments had shown that the application of a Cu substrate for electrodeposition from various non-aqueous media can cause a damage of the Cu substrate. The reason of the damage could stem from the chloride ions present in the metal-chloride solutions as well as the formation of aggressive decomposition products (perhaps radicals) during the solvent reduction. However, Pd-coated Cu foils proved to be a reliable and reproducible substrate. The Pd layer with an approximately 1.4 μm thickness was electrodeposited onto the Cu substrate from alkaline $PdCl_2$ solution. For preparing the Pd coating, $PdCl_2$ (0.05 mol $dm^{-3}$) was dissolved in an $NH_4Cl$ solution (0.1 mol $dm^{-3}$); then, this stock solution was slowly poured under vigorous stirring into 0.5 mol $dm^{-3}$ NaOH solution of 10 times larger volume (for the original recipe see Ref. 20). By using this bath, the spontaneous exchange reaction between Cu and $Pd^{2+}$ was completely eliminated and the process provided a uniform Pd coating. The current density yielding a compact Pd deposit was –0.4 $mAcm^{-2}$.

All chemicals used for the solution preparation were of analytical grade. The Ni samples were electrodeposited from electrolytes containing dry $NiCl_2$ (Alfa Aesar, anhydrous, 99%; 0.1 mol $dm^{-3}$) and also NaCl (Reanal, 0.1 mol $dm^{-3}$) as supporting electrolyte. The non-aqueous solvents: methanol (Reanal), ethanol (Reanal), ethylene-glycol (Alfa Aesar), glycerol (Alfa Aesar), 1,2-propanediol (Alfa Aesar) and 1,3-propanediol (Acros Organics), all of "pure" quality, were used as received without any further purification.

*Sample preparation.* — Electrodeposition was performed at ambient conditions in a simple cell using a 3-electrode configuration. The counter electrode was a Ni sheet immersed in the electrolyte parallel to the vertically aligned cathode, and a saturated calomel electrode (SCE) served as reference. For the preliminary cyclic voltammetry study, a Pt working electrode of 1.09 $cm^2$ surface area was used. In all cases, the scan rate was 5 mV/s. For the preparation of samples used for various further experiments (SEM, XRD or magnetization study), a Pd-coated Cu foil of 20 mm diameter was applied (total geometric surface area 3.14 cm$^2$).

Samples were obtained by d.c. plating under potentiostatic control with an EF453 type potentiostat/galvanostat (Electroflex, Hungary) and with an IviumStat (Ivium Technologies, The Netherlands). The total charge was 10.12 C during the deposition process, so the nominal thickness of the samples was 1 μm by assuming 100% current efficiency. For assessing the actual current efficiency from the alcoholic solutions, a control Ni sample was deposited from a Watts-type aqueous bath under the same conditions. For quartz crystal microbalance (QCM) experiments, an SRS-200 instrument was used with Pt-coated crystals of 5 MHz frequency.

*Sample characterization.* — The sample composition was determined by energy-dispersive X-ray spectroscopy (EDX) in a JEOL JSM 840 scanning electron microscope. The RÖNTEC EDX unit equipped with a Si detector was operated at 25 kV. The chemical analysis was carried out for spots of about 1 mm$^2$ surface area each. The composition was measured on at least three different spots for every sample. The SEM images on the surface morphology were also recorded in this instrument.

XRD patterns with Cu-K$_\alpha$ radiation were recorded on a Philips X'pert powder diffractometer in $\Theta$-$2\Theta$ geometry. The step width was 0.02 degree in the range of 20–100 degree. For a quantitative evaluation of the XRD patterns, each clearly visible peak was fitted with a Gaussian function from which the peak position and the peak width were deduced. For each diffraction line, the corresponding lattice plane distance was determined from the peak position with the Bragg formula. From the line width of the individual XRD peaks, the grain size was estimated by using Scherrer's formula,[21] after correcting the measured linewidth for instrumental broadening derived from the XRD pattern for a LaB$_6$ single crystal. Relative texture coefficients for an *hkl* orientation were calculated with the following formula:

$$RTC(hkl) = \frac{I(hkl)/I_0(hkl)}{\sum_{hkl} I(hkl) \Big/ \sum_{hkl} I_0(hkl)}$$

where $I$ and $I_0$ are the peak intensities observed in the actual and reference powder diffractograms.

The magnetic properties were examined at room temperature by using a vibrating sample magnetometer (VSM). The magnetization data were were recorded in the magnetic field range of $H = \pm 5$ kOe after saturating the samples in a field of $H_s = \pm 12$ kOe.

## Electrochemical characterization of solutions

*Conductivity of the $Ni^{2+}$ solutions.* — The conductivity of the electrolyte solutions used for cyclic voltammetry studies and for the deposition of Ni is shown in Table 1. The viscosity of the solvents used increases by more than three orders of magnitude as the chain length and the number of hydroxyl groups increases from methanol to glycerol. However, the solubility of the Ni salt and using NaCl as supporting electrolyte is sufficient and the decrease in conductivity is moderate as compared to the change in viscosity. Although the conductivity clearly decreases as the viscosity of the solvent increases, the conductivity decreases by about 50 times only while the viscosity increases by a factor of about 2400.

The increase in the conductivity when the $MeCl_2$ salt (Me = Ni or Ca) is added to the 0.1 mol/liter NaCl solution is higher for the nickel salt than for the calcium salt. This trend suggests that the degree of dissociation is higher for $NiCl_2$ than for $CaCl_2$. Since the ion radius of the $Ni^{2+}$ and $Ca^{2+}$ ions are 72 pm and 99 pm, respectively, the higher conductivity of the $Ni^{2+}$-containing electrolyte solutions may also stem from both the difference in specific molar conductivities of the $Me^{2+}$ ions and the difference in the degree of dissociation.

The relatively high conductivity of the solutions made it possible to perform the voltammetry experiments without any special care. Since the distance between the working and reference electrode was about 1 mm, no ohmic drop correction was necessary.

*Polarization characteristics of the Ni deposition and dissolution.* — Figure 1 shows the cyclic voltammograms recorded for the alcoholic solvents (methanol, ethanol, ethylene glycol, 1,2-propanediol, 1,3-propanediol and glycerol). The onset potential of the deposition of Ni was studied in comparison with the dual blank solutions. The blank solutions contained NaCl as supporting electrolyte, and one of them contained $CaCl_2$ instead of $NiCl_2$ in the same concentration as the original Ni-salt component of the plating bath. The polarization behavior of Ni deposition from various alcoholic solvents is not known from the literature.

The deposition of Ni starts at different potentials for the various alcoholic solvents, ranging from −0.75 V to −1.00 V. A nucleation barrier can be observed in some solvents (methanol and 1,2-propanediol), which manifests itself in the difference between the onset potential of metal deposition in the cathodic-going curve and the potential at which the metal deposition ends in the anodic-going scan (the latter being more positive). This causes an opposite order of the cathodic-going and anodic-going curves (anomalous hysteresis) and a crossing in the voltammograms.

On the anodic-going scans, the end of the Ni deposition can be observed nearly at –0.75 V in all solvents used. The cathodic current at $E > -0.75$ V in the cathodic-going scans (i.e., prior to the onset of the deposition) can be attributed to the reduction of the impurities in the solution. Several attempts were made for the removal of oxygen by bubbling the solutions with argon, but this did not prove to be as efficient as for aqueous solutions. After one hour of Ar bubbling, the reduction of the cathodic current at $E > -0.75$ V was at most 50%, but the deaeration could not diminish the background current observed for the blank solutions at $E < -1.0$ V. Hence, it is concluded that at potentials more negative than –1.0 V, the decomposition of the solvent also takes place.

The Ni deposition leads to an increase in cathodic current as compared to the same solution with no $NiCl_2$. In some cases, however, the increase in the cathodic current density is negligible when a comparison is made with the solution containing $CaCl_2$ instead of $NiCl_2$. This finding draws the attention to the fact that the solvent decomposition, which is the cathodic reaction in the absence of $NiCl_2$, remains significant when the Ni deposition also takes place in a parallel reaction.

The onset potential of the Ni dissolution strongly depends on the solvent used. The most positive onset potential of Ni dissolution was found for methanol where the Ni dissolution started at –0.1 V. At small deposit thickness, a single dissolution peak was observed in methanol, which often splits up into a pair of peaks at high deposit thickness. The reason for the occurrence of this feature of the voltammograms could not be explained.

The process characterization given above was confirmed for methanolic solutions by using a quartz crystal microbalance. (The viscosity of other solvents leading to metalic deposit was too high to operate the microbalance in the resonating mode, and this prevented us from similar observations with other alcoholic solvents.) The derivative of the QCM data can be seen in the top part of Fig. 1(a). This confirmed that no deposition process takes place in the potential interval of $-0.25$ V $> E > -0.75$ V, although a significant cathodic current was observed in this potential range. This supports the assumption that the cathodic current in this potential range can be attributed to either the reduction of the dissolved oxygen or the cathodic decomposition of the solvent (or both). Metal deposition can take place at potentials $E < -0.75$ V. The QCM data show the same type of hysteresis due to the nucleation barrier as the voltammetric data. The apparent molar weight of the deposit dissolving in methanol during the anodic process was also calculated. The weight loss of the electrode and the charge passed in the anodic peak resulted in the molar weight of the deposit dissolved and was found to be 55.9 gmol$^{-1}$. This compares well to the 58.7 gmol$^{-1}$ molar weight of Ni within the sensitivity of the QCM. Since the coating thickness at the edge of the QCM crystal (where the differential sensitivity is lower than at the center) can be somewhat higher due to the larger accessibility for the reactant, the small negative deviation form the theoretical molar weight is perfectly understood.

When the conductivity of the solution is fairly low (higher-valence alcohols, especially glycerol), the dissolution peak is rather wide. This can be explained by the hindrance of the transport processes. This slow ionic transport in the solution as a reaction step following the charge transfer process may also become a limiting factor of the dissolution rate.

The current efficiency was determined from the ratio of the anodic and the cathodic charge passed during the cyclic voltammetry measurements for all solutions. When the cathodic limit of the CV was −1.5 V, the lowest current efficiency was obtained for ethanol, namely 29 %. In the case of glycerol, the largest current efficiency (76 %) was observed. In the case of other alcohols, this value was between 61 and 70 %. For less negative cathodic limits, the current efficiency of the Ni deposition was significantly smaller (typically, at −1.0 V the current efficiency was about half of the value measured at −1.5 V). It is to be stressed that the current efficiency calculated with the above mentioned method comprises the entire anodic charge measured without distinguishing the metal dissolution and the reoxidation of any other product forming on the cathode during the negative-going scan. Therefore, the current efficiency data are indicative of the upper limit of the current efficiency of the metal deposition.

## Morphology, composition and structure of deposits

Samples were deposited from various alcoholic solvents in potentiostatic mode for the study of the surface morphology, composition and crystalline structure. In most of the cases, morphology and structural data refer to the same specimens.

*Ni deposits obtained from methanol.* — Ni deposits with metallic appearance could be grown at various constant deposition potentials in the range −0.9 V to −1.4 V from the methanolic bath. For deposition potentials more negative than −1.40 V, the thickness of the nickel deposits was apparently much smaller in comparison with deposits at more positive potentials with the same charge passing the cell. This sudden change beyond $E = -1.4$ V cannot be explained with the current efficiency data obtained from the cyclic voltammograms. The strong decrease in the deposit thickness at these highly negative cathodic deposition potentials may be due to a loss of adhesion of the deposit already during the deposition process because of the residual stress.

Figure 2 shows scanning electron micrographs at two magnifications for samples prepared with methanolic solution in the potential range –0.9 V to –1.4 V. At –0.90 V deposition potential, the deposit has a fragmented morphology as indicated by the fracture lines. Compact Ni deposits could be obtained in the potential range of –1.1 to –1.4 V. The disappearance of the stress-induced fractures was accompanied by the formation of a granular morphology with a surface granule size in the submicrometer range. These coatings were unambiguously continuous on the surface.

The EDX analysis revealed that the deposit consists of Ni. Besides Ni, only Cu was found as a major constituent, and occasionally Pd could also be seen in the spectra, both latter elements belonging to the substrate.

Typical X-ray diffractograms for samples deposited from methanol solution can be seen in Fig. 3. The sample prepared at –0.90 V deposition potential differs from the rest, while the diffractogram shown for the sample obtained at –1.25 V is characteristic also for all other samples. At low deposition potential (–0.90 V), only a weak line could be identified in the diffractogram that may correspond to the Ni (111) reflection. The common features of the diffractograms of deposits made from methanol in the deposition potential range –1.0 V to –1.4 V can be summarized as follows. The diffractograms are dominated by diffraction lines that belong to fcc-Ni, accompanied by lines characteristic of the fcc-Cu substrate. All diffraction lines characteristic of the stable fcc crystal structure of Ni can be easily identified in the diffractogram in the $2\Theta$-range scanned. There is no significant difference in the texture of the Ni samples deposited from methanol at potentials in the range –1.0 V to –1.4 V. The (111) reflection is stronger than expected for a random crystal orientation and its relative texture coefficient is as high as 1.58. The other peaks corresponding to the (200), (220) and (311) reflections exhibit successively a lower texture coefficient of 0.97, 0.70 and 0.42, respectively. Hence, the diffraction pattern indicates a weak (111) texture.

For the deposit obtained at –0.9 V, due to the single weak fcc-Ni reflection, the lattice constant could not be reliably evaluated. For the rest of the samples, the lattice constant was 0.3525 nm on the average which is fairly close to the value 0.35232 nm reported for pure fcc-Ni (Ref. 22).

From the line width of the individual XRD peaks, the grain size was estimated by using Scherrer's formula.[21] This calculation yielded uniform values for all samples and diffraction lines with an average of 10 nm. The use of Scherrer's formula usually underestimates the grain size to some extent due to neglecting any other source of line broadening (like the microstrains); thus, the actual grain sizes can be somewhat larger than those calculated.

*Deposits obtained from ethanol.* — Electrodeposited samples were prepared at −1.10 V and −1.40 V deposition potentials from ethanol. The SEM picture of the deposit prepared at −1.40 V is shown in Fig. 4. The long linear pattern in the fragmentation of the deposit follows the rolling direction of the substrate. Although the fragmented morphology of the samples appears to be similar to the sample made from methanol at −0.90 V deposition potential, clear differences can also be seen. The surface morphology of the deposits obtained from ethanol at the more negative potentials (−1.10 V and −1.40 V) is not similar to the cauliflower-like deposit morphology of metallic Ni obtained from methanol at the corresponding potentials (Fig. 2).

The EDX analysis revealed that the material accumulation on the substrate contains Ni. In addition, Cu and occasionally Pd also appear in the EDX spectra as substrate components. Evidently, the relative intensity of Ni with respect to the Cu and Pd signals depends on the deposit thickness. For example, the intensity of Ni in the deposit obtained at −1.10 V indicated a relatively thick Ni-containing layer, i.e., the Cu signal was fairly weak whereas at −1.40 V, the amount of Cu detected was almost comparable to that of Ni, i.e., the Ni-containing layer was not very thick. The latter finding is in agreement with the estimate from the CV curves in a previous section in that the current efficiency was found to be very small (29 %) in ethanol for large cathodic limit (−1.50 V).

It should be added, furthermore, that the EDX analysis also revealed in some cases a significant amount of the non-metallic elements (O and Cl) in the Ni deposits from the ethanol bath. A quantitative evaluation of Cl yielded an atomic ratio of Ni:Cl ~ 75:25. Such an amount of chlorine in the deposit may stem, most probably, from the presence of $NiCl_2$ salt precipitates in the deposit. The concentration of O could not be quantitatively assessed due to mainly a baseline problem but according to a rough estimate the atomic fraction of O in the deposit was comparable to that of Ni at −1.0 V deposition potential. The analysis data thus indicate that either NiO may be present in the deposit or, alternatively, some NiClX compound may form where X refers to an anion produced in the cathodic reaction (e.g., methanolate).

According to the XRD measurements, for most samples produced from the ethanolic bath, no fcc-Ni lines could be identified in the diffractograms which were dominated by the lines characteristic of the fcc-Cu substrate. Some low-intensity diffraction peaks were also present which can eventually be ascribed to the compounds NiO and $NiCl_2$. In agreement with the qualitative conclusion drawn from the EDX data, the above XRD results clearly exclude the formation of metallic nickel. The absence of ferromagnetic Ni in the deposit is also in accord with the results of magnetic measurements to be described later. However, photoelectron spectroscopy data are necessary to unambiguously reveal the chemical bond state of Ni, Cl and O and, thus, to confirm the eventual presence of $NiCl_2$, NiO and/or other organic nickel salts in the deposit.

*Deposits obtained from ethylene glycol.* — Samples were prepared at various deposition potentials (from −1.15 V to −1.35 V) from ethylene glycol solution. Regardless of the deposition potential, the deposit exhibited a lamellar structure. A SEM picture of the samples obtained from ethylene glycol can be seen in Fig. 5. The morphological transition seen for methanolic solution as a function of the deposit potential could not be seen for ethylene glycol. The size of the surface features was nearly 1 μm and 10 μm for samples deposited at −1.20 V and −1.35 V, respectively. The size of the surface features was smaller than for a deposit obtained from ethanol where the distance between the fracture lines was a few tens of micrometers (see Fig. 4).

The EDX analysis revealed that the deposits contain Ni. Besides the dominating Ni component, the presence of Cu due to the substrate could be detected, and occasionally, Pd was also seen in the spectra. The spectra indicated the presence of some oxygen in the deposits which was estimated to be about 5 at.% with respect to Ni.

In the XRD pattern of the sample prepared at the least negative deposition potential (−1.15 V), only the diffraction lines of the substrate components can be identified.

At −1.20 V, −1.30 V and −1.35 V deposition potentials, in addition to the lines characteristic of the fcc-Cu and fcc-Pd substrate, several fcc-Ni lines is observed in the diffractograms in the 2$\Theta$-range scanned. The relative intensity of the (111) reflection is stronger than that expected for a random crystal orientation. The (200) and (220) reflections are much weaker, while the (311) reflection is missing. The intensity of the minor peaks was too small for a quantitative evaluation of the texture coefficient. The lattice parameters derived from the peak positions were close to the value reported for pure Ni (Ref. 22) but a precise quantitative agreement was not expected due to the low intensities of the Ni peaks which, furthermore, strongly overlapped with the high-intensity Cu peaks of the substrate. The Ni grain size estimated by using Scherrer's formula[21] from the line width of the XRD peaks was about 20 nm.

*Deposits obtained from 1,2-propanediol.* — Samples were prepared at −1.30 V and −1.40 V deposition potential from 1,2-propanediol solution. The SEM images (not shown) indicate that only a thin film could be produced on the substrate. The film thickness was so small that the surface pattern of the substrate dominated the morphological character of the deposit.

The results of both the EDX analysis and the X-ray diffraction are similar to the results of the samples which were made from ethanol solution. Namely, the EDX analysis revealed that the samples contain nickel but weak diffraction lines characteristic of fcc-Ni could only be identified on the diffractograms.

*Ni deposits obtained from 1,3-propanediol.* — A SEM picture and an EDX spectrum of the sample prepared at −1.40 V from 1,3-propanediol can be seen in Fig. 6. The EDX analysis shows a significant Ni content in the deposit. A trace amount of chlorine can be identified apart from Ni whereas the O signal is negligible.

The fcc-Ni reflections (111), (200), (220), (311) and (222) could be identified in the X-ray diffractogram, and fcc-Pd and fcc-Cu lines originating from the substrate could also be seen. The relative intensity of the (111) reflection is stronger than in the case of a random crystal orientation, and it is stronger than in the case of methanol and ethylene glycol solutions. The (200), (220) and the (311) reflections are much weaker than in the case of a random crystal orientation. For a deposit obtained at −1.40 V, the lattice constant was fairly close to the value reported for pure Ni (Ref. 22).

*Ni deposits obtained from glycerol.* — Samples were prepared at deposition potentials from −1.00 to −1.40 V from glycerol. The deposit morphology was characteristic of salt-like films at every deposition potential tested. However, there is a clear difference in the sample morphology as a function of the deposition potential. When the film formed at low cathodic polarization (e.g., between –1.00 and –1.15 V), the distribution of the lamellae was random and independent of the surface morphology of the substrate. At higher cathodic potential, however, the deposit structure followed the rolling pattern of the substrate. A SEM picture of two samples obtained from glycerol at two deposition potentials can be seen in Fig. 7.

The EDX analysis revealed that the samples do not contain Ni. The XRD data did not show any indication of the presence of metallic Ni. This is in agreement with the lack of a ferromagnetic behavior of the deposits from glycerol bath to be described in the next section.

## Magnetic properties

The room-temperature magnetization curves were measured on sample pieces of a few millimeters in size. The $M(H)$ curves were qualitatively very similar to each other for all Ni deposits obtained from the various non-aqueous solutions and at various deposition potentials. Furthermore, they were also similar to that measured for the reference Ni sample electrodeposited from an aqueous solution and even the coercive field and remanence values showed a reasonable agreement. A representative $M(H)$ curve measured for a Ni foil deposited from methanolic solvent with a potential of –1.0 V vs. SCE is shown in Fig. 8a.

The similarity of the $M(H)$ curves for Ni deposited from methanol solvent with potential values between –0.9 and –1.4 V vs. SCE is demonstrated by displaying them in Fig. 8b in the low-field range after normalizing the magnetization for the saturation values. A similar plot is shown in Fig. 9 for the $M(H)$ curves of Ni foils deposited from various other alcoholic solvents (ethylene glycol, 1,2-propanediol, 1,3-propanediol) at several potentials.

It has to be mentioned that deposits obtained from ethanol and glycerol contained no ferromagnetic material and the magnetization curves showed a weak diamagnetic baseline, corresponding to the Cu substrate. The results in these latter samples, at the same time, also indicated that the magnetization measured for the other samples cannot stem from the impurity of the substrate.

The coercive field ($H_c$) and relative remanence ($M_r/M_s$) values of Ni foils deposited from the non-aqueous electrolytes as well as of the reference Ni foil from the aqueous electrolyte are collected in Table 2.

As for the Ni foils obtained from the methanolic solution, the deposition potential apparently does not influence significantly the bulk magnetic properties of the deposits. The relative remanence values ($M_r/M_s$) are scattered around an average of 0.5 with values between 0.39 and 0.63. The coercive field values scattered from 131 Oe to 191 Oe, without any apparent dependence on the deposition potential.

In comparison with the Ni samples deposited from methanolic bath, the relative remanence of the Ni foils obtained from the other alcoholic solutions, which exhibited ferromagnetism at all, is somewhat higher with values ranging from 0.52 to 0.71. The coercive field values are also definitely higher here and range from about 226 to 251 Oe, without any clear dependence on solvent or deposition potential. All these parameters indicate that the magnetic hysteresis loops for the other alcoholic solvents was more rectangular than for Ni deposits from the methanolic bath.

These results indicate that ferromagnetic Ni metal can be successfully electrodeposited from various alcoholic solvents and the resulting deposits show very similar magnetic behavior and properties as Ni metal when deposited from the conventional Watts type aqueous bath. In cases where no ferromagnetic signal was detected, it may have been due to the extreme low amount of metallic pure Ni deposited or, in samples with high O content, due to the formation of the NiO compound which is antiferromagnetic at room temperature and it does not give, therefore, a signal in the magnetic measurement.

## Discussion

Although the prospective benefits of the non-aqueous solvents have already been recognized, this was not followed by a comprehensive study of the possible solvents as electrodeposition media but previous investigations have rather focused on the application of one solvent of choice only. Furthermore, in spite of the numerous reports published so far on electrodeposition of metals from non-aqueous solutions, little is known about the details of the electrodeposition of metals in such media since most of the studies have been dealing with the deposit quality rather than the fundamentals of the processes.

A comparative study of nickel electrodeposition was reported in the present paper by using a number of alcoholic solvents. Besides methanol and ethanol as solvents, Ni electrodeposition from ethylene glycol, glycerol, 1,2-propanediol and 1,3-propanediol was also the aim of our research. As to the source of $Ni^{2+}$ ions, we have chosen nickel chloride as a commercially available and inexpensive solute in the alcoholic solvents. In some studies known from the literature, the influence of various solution components is difficult to separate. Lee at al.[14] used a $NiCl_2 \cdot 4H_2O$ as solute; therefore, water was present in spite of all efforts to produce Ni from a non-aqueous bath. Similarly, Shrestha and Saji[15] mixed ethanol with both $NiCl_2 \cdot 6H_2O$ and aqueous HCl solution to obtain Ni deposits. It is also difficult to assess the impact of the trifuoroacetate anion on the deposition process when various ammonium and alcali metal halogenides were applied in the same bath.[5] As compared to the above mentioned works, the present study offers an opportunity of the clear comparison of alcoholic solvents in the absence of water.

A detailed cyclic voltammetry study of these solutions has been carried out by establishing the polarization characteristics of Ni deposition and dissolution in each bath. The Ni deposition potential could be established for most of the solvents, but no electrodeposition of Ni was seen in the voltammograms of glycerol, which was confirmed by XRD, EDX and magnetization measurements, too.

Nickel deposits obtained from solvents other than methanol were of lower quality in most cases. The Ni content in the deposits was in some cases fairly low and the characteristic fcc-Ni lines could also not always be observed. This might have been due to a low current efficiency in these cases or due to a peeling off of the deposits from the substrate as a consequence of large internal stresses accumulated during deposition. In some deposits, we could observe chlorine present, most probably due to $NiCl_2$ salt precipitation. Some deposits contained also a significant amount of O to the extent that the formation of NiO compound may have also occurred as hinted at also by some weak XRD lines in the diffractograms.

A discrepancy was found between the results of the voltammograms and the deposit features in some solvents, especially for ethanol. While the increase in the cathodic current in the Ni deposition potential range and the occurrence of a corresponding anodic peak can be seen in the voltammograms, neither can any Ni deposit be identified with XRD nor ferromagnetic behavior is observed in the magnetization study. At the same time, the deposit morphology is characteristic of a salt film rather than a metal layer. This behavior was associated with a salt film formation.

Little is known about the cathodic reactions of alcoholic compounds. From the earliest time of the investigation of this topic, the only cathodic reaction assumed was the reductive deprotonation of the –OH group, accompanied with hydrogen evolution (R–OH + e = R–O$^-$ + ½ H$_2$).[23] The most recent source of information on the cathodic behavior of alcohols also confirms[24] that no other oxygen-containing organic compounds can form during the cathodic treatment of alcohols and their reduction to alcanes can be performed after activation.

While the cathodic behavior of alcohols generated a very modest interest only and attention was rather paid to the oxidation of alcohols in fuel cells, methanol became an important component of low-voltage hydrogen generation systems.[25-27] Such electrolysis cells are operated with acidic solutions of methanol, and the cathode reaction is the hydrogen evolution, the energy gain being provided by the low electrode potential of methanol oxidation as compared to water splitting. Studies performed in this field showed that practically the only cathode reaction is the hydrogen evolution, and impurities observed in the cathode compartment of the cell is due to the incomplete separation of the anode.[27] Based on all these observations, we also assume that the only side reaction at the cathode during the nickel deposition is the hydrogen evolution with the formation of alcoholate anions.

The non-metallic deposits formed in some of the alcoholic solutions (especially in ethanol and glycerol) are assumed to be nickel(II) alcoholates and mixed alcoholates and chlorides. The diffractograms did not indicate the formation of a phase with definite crystal structure. This is not surprising, taking also into account that the solvent content of these deposits may also vary and they can be amorphous.

It is to be noted that a significant difference may arise in the deposit nature as the electrolysis proceeds. The formation of a film can start when the vicinity of the cathode becomes supersaturated with respect to a compound. Therefore, it is possible that a thin Ni film is formed on the cathode before the precipitate can form. This is why pronounced anodic peak can be observed also in ethanol where no trace of metallic nickel can be seen after a prolonged electrolysis.

The room-temperature magnetization measurements revealed a ferromagnetic behavior in most deposits except those obtained from ethanol and glycerol. The magnetic hysteresis loops were qualitatively similar for all ferromagnetic samples. The coercive field varied from 130 to 250 Oe, with somewhat lower values for the deposits from the methanolic solution whereas the relative remanence was in the range 0.4 to 0.7, for the methanolic solution exhibiting again somewhat smaller values. The variation of the magnetic parameters showed no systematic dependence on either the deposition potential or solvent except for the difference between values for the methanolic bath and the other solvents.

A structural and magnetization study of Ni deposited from methanolic bath (although made with nickel salt containing water) was given by Lee et al.[14] Although the pulse

- 16 -

deposition method and the application of a Si substrate could not lead to a uniform deposit in the latter work, the correlation of the magnetic behavior with the grain size could be established. The coercive field of Ni varied between 60 and 210 Oe,[14] which is comparable to the results of the present work. Although no detailed evaluation of the diffractograms was offered by Lee et al.,[14] it can be assessed that their deposits showed a stronger (111) orientation than the samples prepared from methanol in the present study.

As a summary, it was shown that Ni deposits can be successfully obtained from alcoholic solutions, especially from methanol This may have more far-reaching impact on the electrochemical deposition of nickel-based alloys with such alloying element that are not compatible with any aqueous bath due to their sensitivity to water or their very negative deposition potential.

## Conclusions

1. It has been demsontrated that dry nickel(II) chloride can be used for the preparation of alcohol-based electrolyte solutions. The $Ni^{2+}$ concentration can be as high as 0.1 mol dm$^{-3}$, and NaCl can also be used as a supporting electrolyte in the same concentration. The conductivity of the solutions is sufficient for electrochemical experiments and it decreases as the viscosity of the solvent increases.

2. From the alcohols investigated, methanol proved to be an appropriate medium for Ni deposition from dry $NiCl_2$ salt. Compact Ni deposits were obtained in the deposition potential range of −1.10 V to −1.40 V vs. SCE. The XRD study of these deposits revealed that crystal structure corresponds well to pure Ni concerning the lattice constant, while a weak (111) texture can be established.

3. The product of the anodic dissolution process in methanol was investigated with quartz crystal microbalance. The molar weight of the component dissolved was 55.9 gmol$^{-1}$, which is in fair agreement with the molar weight of Ni (58.7 gmol$^{-1}$).

4. Deposits containing Ni were obtained from ethylene glycole, 1,2-propanediole and 1,3-propanediole. However, no metallic Ni was found in the deposits obtained from ethanol and glycerol. In the latter cases, a salt film was obtained on the substrate, which was a result of the chemical precipitation of the alcholate anions and nickel cations. No ferromagnetic component was found in the salt film deposits.

5. Where metallic Ni was present in the deposit, the magnetic properties of the film was comparable to those of Ni plated from a conventional aqueous bath, regarding to both squareness of the magnetization curves and coercive fields.

## Acknowledgements

This work was supported by the Hungarian Scientific Research Fund (OTKA) through Grant # K 104696.

Table 1: Viscosity of the solvents and conductivity of the alcoholic solutions used for cyclic voltammetry studies and for Ni deposition at 20 $^{o}$C.

| Name of alcohol used | Viscosity of the solvent (Pa s) | Conductivity of the 0.1 M NaCl solution (mS·cm$^{-1}$) | Conductivity of the 0.1 M CaCl$_2$ + 0.1 M NaCl solution (mS·cm$^{-1}$) | Conductivity of the 0.1 M NiCl$_2$ + 0.1 M NaCl solution(mS·cm$^{-1}$) |
|---|---|---|---|---|
| methanol | 0.00059 | 0.538 | 0.755 | 0.769 |
| ethanol | 0.00114 | 0.103 | 0.185 | 0.194 |
| ethylene-glycol | 0.0161 | 0.439 | 0.610 | 0.638 |
| 1,2-propanediol | 0.042 | 0.226 | 0.409 | 0.531 |
| 1,3-propanediol | 0.052 | 0.283 | 0.446 | 0.510 |
| glycerol | 1.410 | 0.0399 | 0.0617 | 0.0761 |

Table 2: Magnetic properties (coercive field $H_c$ and relative remanence $M_r/M_s$) of Ni deposits obtained from various alcoholic solvents and from a Watts type aqueous bath.

| Solvent/bath | Deposition potential vs. SCE (V)vs | $H_c$ (Oe) | $M_r/M_s$ |
|---|---|---|---|
| methanol | −0.90 | 191 | 0.52 |
| methanol | −1.00 | 186 | 0.44 |
| methanol | −1.10 | 131 | 0.39 |
| methanol | −1.25 | 179 | 0.63 |
| methanol | −1.40 | 160 | 0.55 |
| 1,2-propanediol | −1.30 | 233 | 0.71 |
| 1,3-propanediol | −1.30 | 226 | 0.52 |
| ethylene glycol | −1.20 | 251 | 0.54 |
| ethylene glycol | −1.30 | 233 | 0.66 |
| ethylene glycol | −1.35 | 242 | 0.59 |
| Watts bath | | 183 | 0.49 |

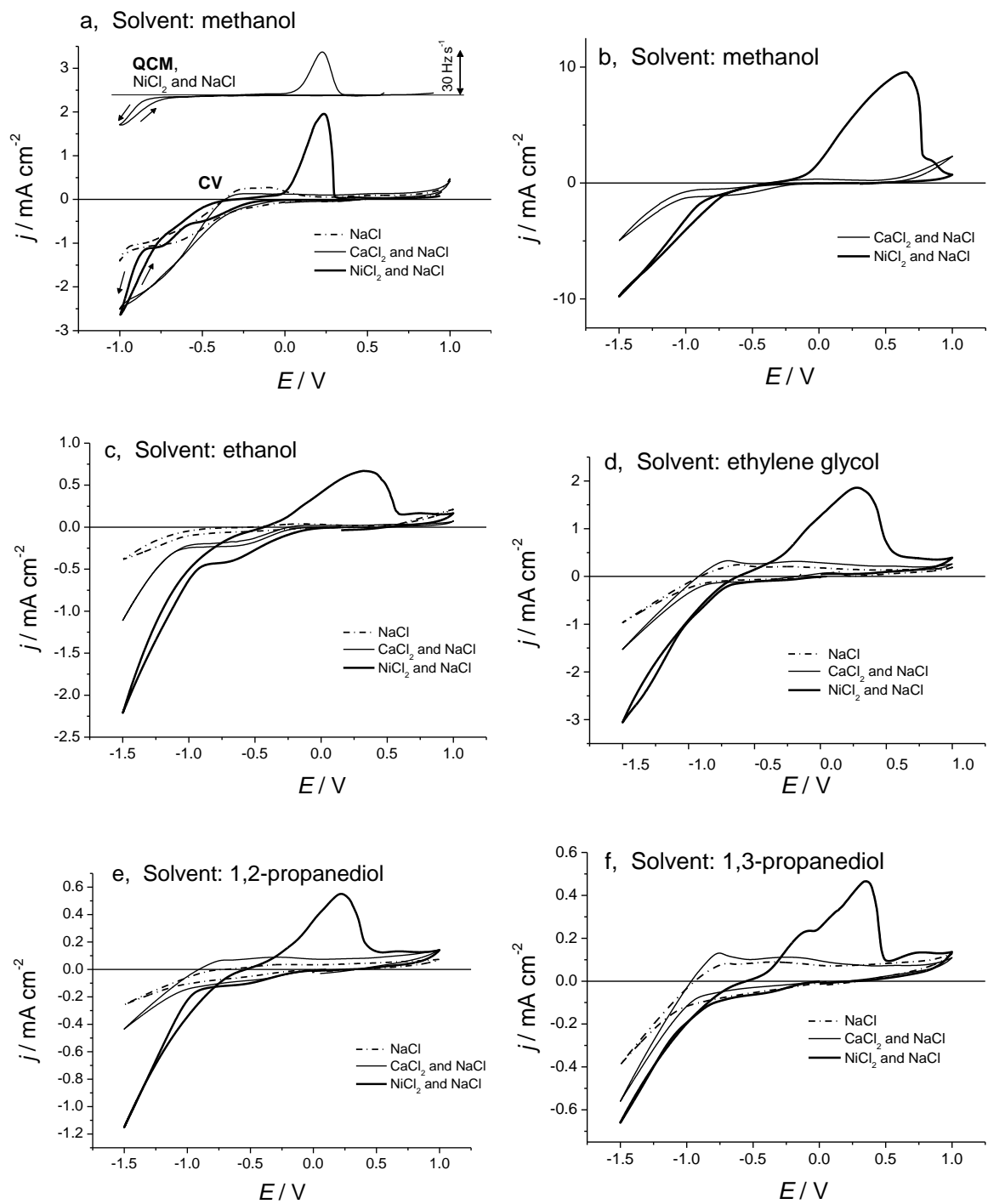

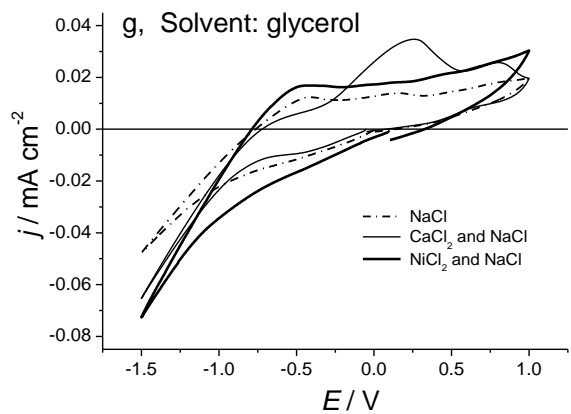

*Fig. 1 Cyclic voltammograms measured on a Pt electrode for various alcoholic solutions containing supporting electrolytes only and NiCl$_2$. The top of Fig. 1(a) presents also the derivative QCM data measured in parallel to Ni deposition, and the arrows indicate the scan direction.*

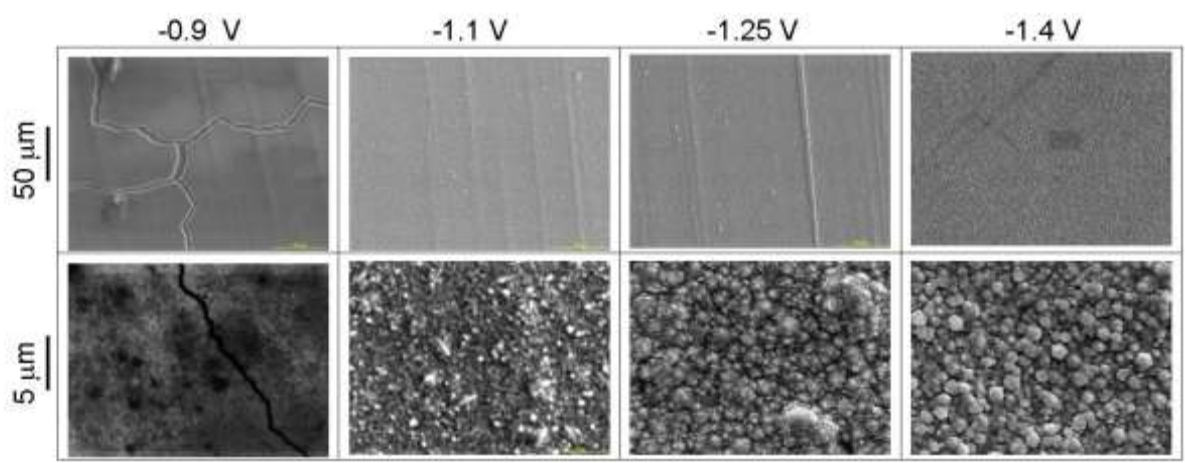

*Fig. 2 Scanning electron micrographs of the deposits obtained at different potentials from methanol.*

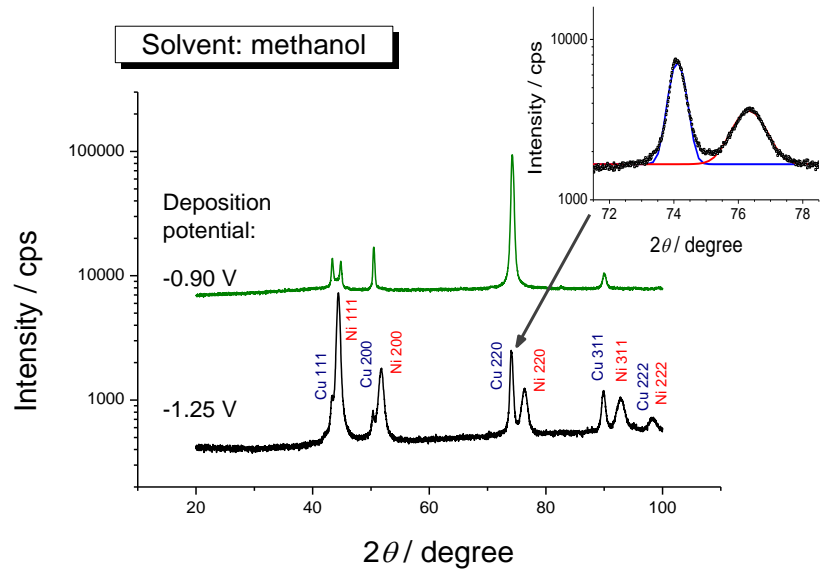

*Fig. 3  Typical X-ray diffractograms of d.c. plated Ni deposits prepared from the methanol bath at -0.90 and –1.25 V. The inset shows the fitting quality of the Gaussian functions used for the correct identification of overlapping peaks.*

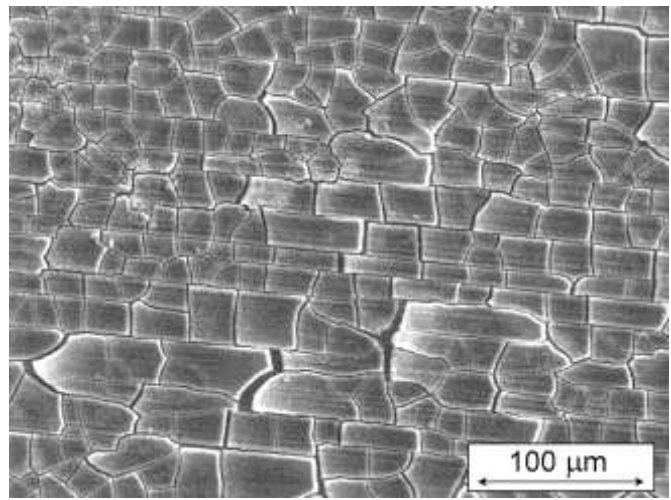

*Fig. 4  Scanning electron micrograph of the deposit obtained at –1.40 V potential from ethanol.*

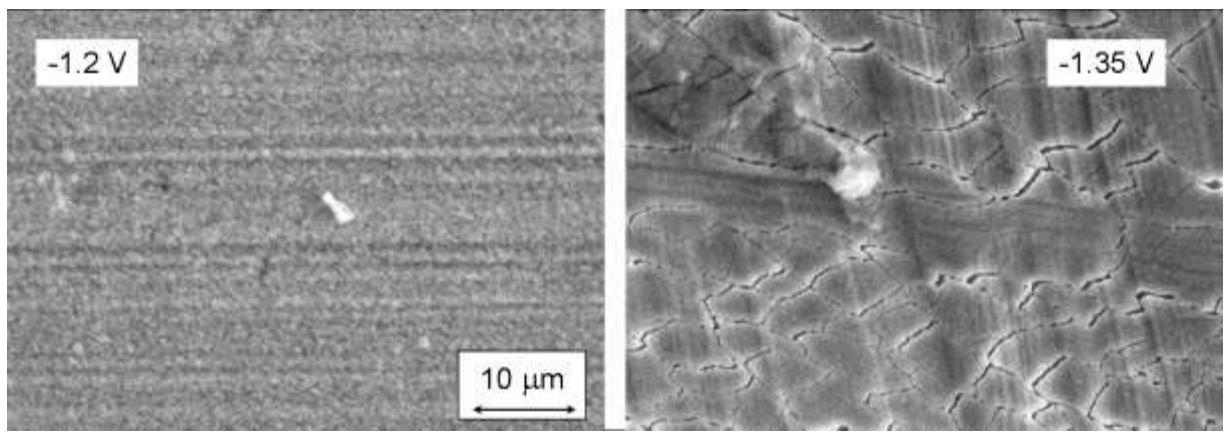

*Fig. 5   Scanning electron micrographs of deposits obtained from ethylene glycol.*

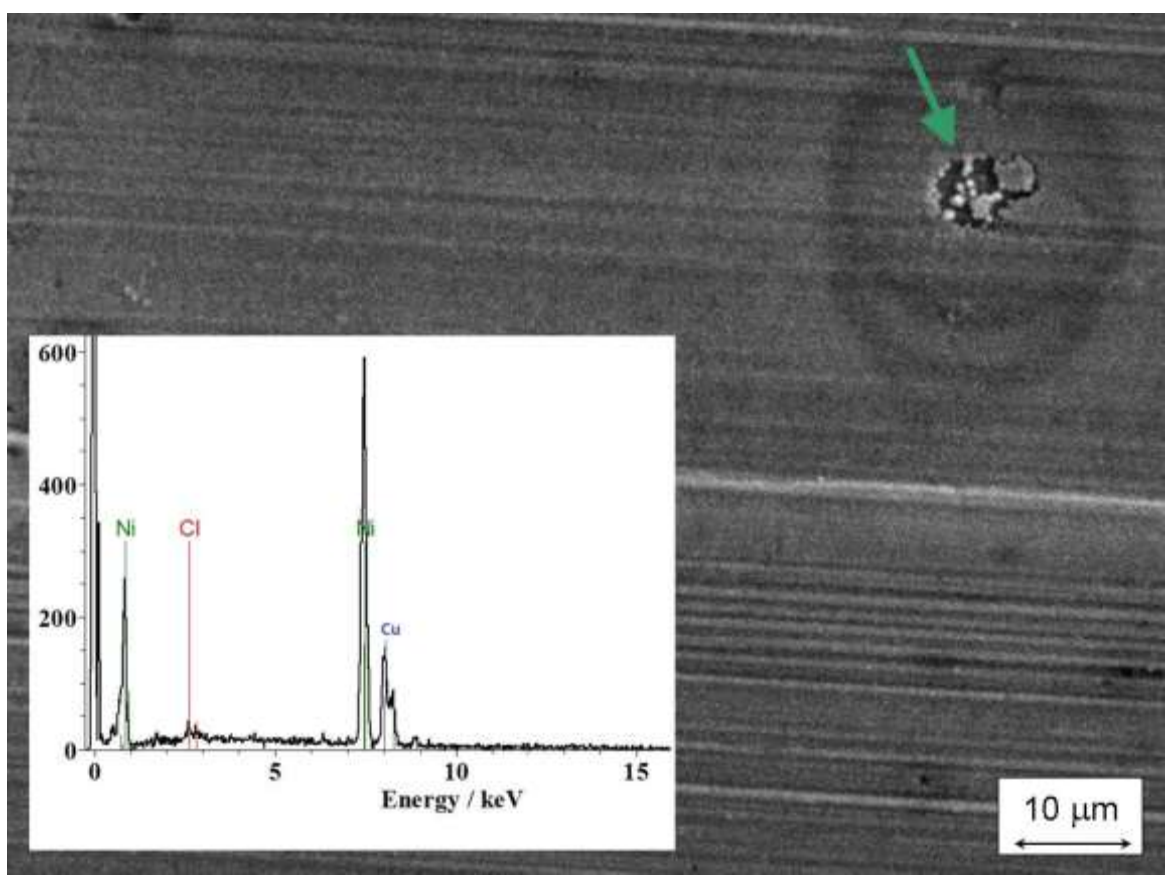

*Fig. 6   Scanning electron micrograph and EDX spectrum of a deposit obtained at –1.30 V from 1,3-propanediol. The arrow shows a defect in the deposit where the presence of the coating can be visually detected.*

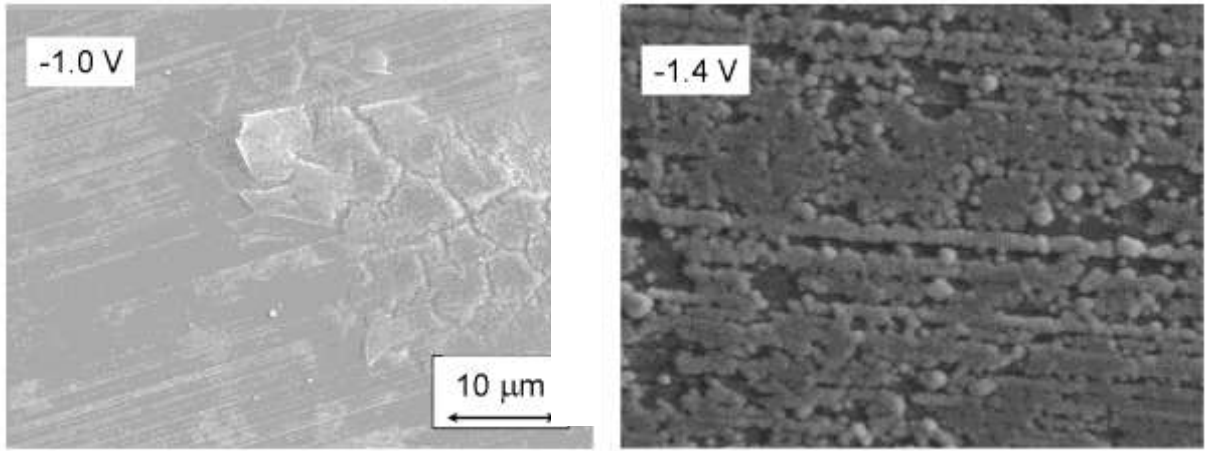

*Fig. 7   Scanning electron micrographs of deposits obtained from glycerol.*

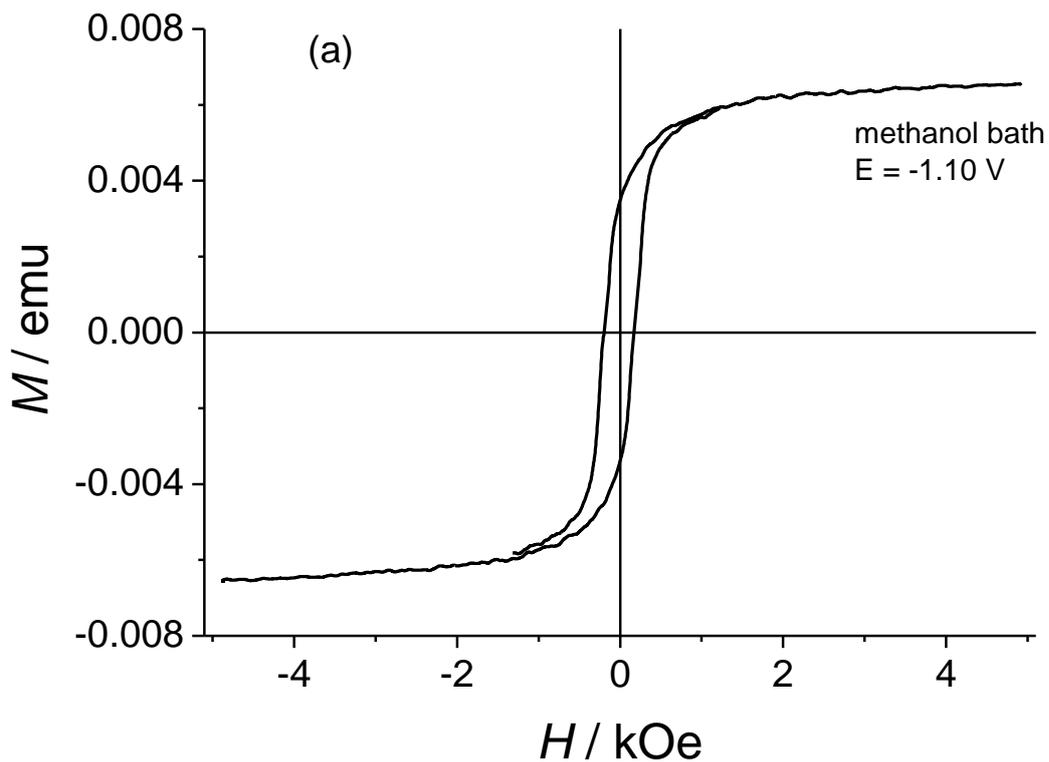

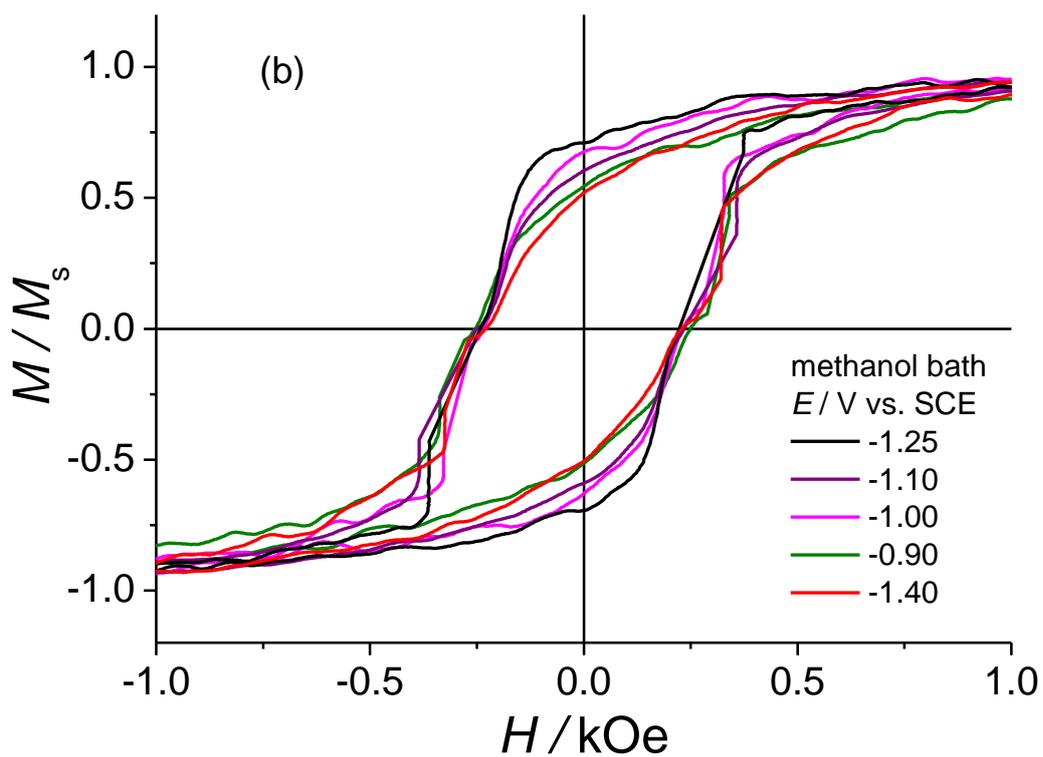

*Fig. 8 (a) M(H) curve of a Ni film electrodeposited from methanol at an electrode potential of -1.10 V vs. SCE. (b) Normalized low-field M(H) curves of Ni films electrodeposited from methanol at different electrode potentials as indicated in the legend.*

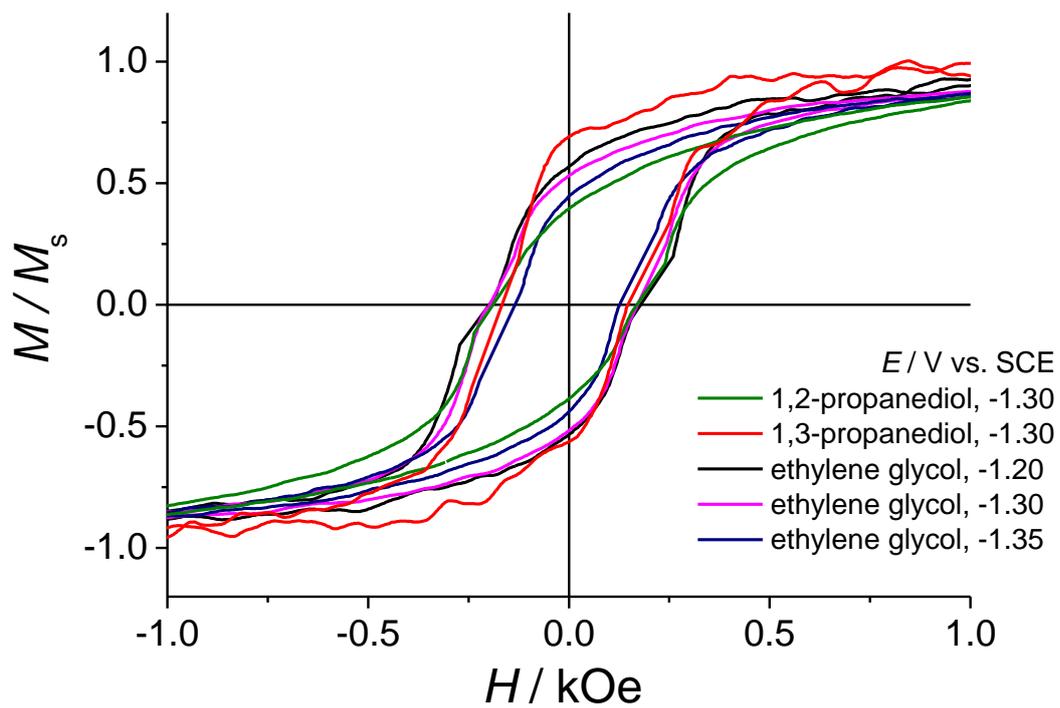

*Fig. 9 Normalized low-field M(H) curves of Ni films electrodeposited from various non-aqueous solvents at different electrode potentials as indicated in the legend.*